\documentclass[11pt,a4]{article}
\usepackage{moriond,epsfig,amsmath,amssymb,psfrag}

\def\be{\begin{equation}}
\def\ee{\end{equation}}
\def\bea{\begin{eqnarray}}
\def\eea{\end{eqnarray}}

\begin{document}
\vspace*{4cm}
\title{IMPROVED BOUNDS ON UNIVERSAL EXTRA DIMENSIONS}

\author{ T. Flacke }

\address{Rudolf Peierls Centre for Theoretical Physics, University of Oxford,1 Keble Road, Oxford OX1 3NP,UK}

\maketitle\abstracts{We report on recent constraints on models with a flat ``universal'' extra 
dimension in which all Standard Model fields propagate in the bulk. A significantly improved 
constraint on the compactification scale is obtained from the extended set of electroweak precision 
observables accurately measured at LEP1 and LEP2. We find a lower bound of $M_c \equiv R^{-1} > 700$ (800) GeV
at the 99\% (95\%) confidence level. Comparison of this constraint with the relic density of Kaluza-Klein dark 
matter for the Minimal UED model points towards the necessity of including non-minimal boundary terms which 
motivates studying alternative Kaluza-Klein dark matter candidates. Results for the one-loop induced magnetic dipole moment 
for Kaluza-Klein neutrino dark matter are presented.\\
This talk is based on  Phys.Rev.D73:095002,2006 and hep-ph/0601161. 
}

\section{Introduction}

Models with flat ``universal'' extra dimensions (UED) in which all
fields propagate in the extra dimensional bulk are of
phenomenological interest for two primary reasons.  First, the mass scale of the compactification is only 
constrained to $1/R\gtrsim 300$ GeV and well within the reach of future collider experiments.  Moreover the collider signatures of Kaluza-Klein (KK) particle production in UED models are easily confused with those of superpartner production in supersymmetric models.
Second, UED models provide a viable dark matter candidate -- the lightest
Kaluza-Klein particle (LKP) -- which is stable by virtue of a
conserved discrete quantum number intrinsic to the model. 

In the next section we give a brief review on UED models before reporting on an improved constraint on $1/R$ , obtained from electroweak precision data of LEP1 and LEP2 in section \ref{s2}.\cite{flacke2} Comparison with constraints on KK dark matter points towards the necessity of including non-minimal boundary terms. Their inclusion can change the LKP, which motivates our investigation of potential bounds on the KK neutrino. In section \ref{s3} we report on the magnetic dipole moment of the Kaluza-Klein neutrino, induced at first loop order.\cite{flacke3} 

\section{Universal extra dimensions}
In UED models, all Standard Model (SM) fields are promoted to 5D fields. In order to obtain chiral zero modes for the fermions which are to be
identified with the Standard Model fermions, the model is
compactified on $S^1/Z_2$. By integrating out the extra dimension, 
every 5D field yields an infinite tower of effective 4D Kaluza-Klein 
modes, which, in the case of fermions, are Dirac fermions for all
but the chiral zero mode. The mass spectrum of the $n^{\mbox{th}}$ KK modes is to good
approximation an SM spectrum, shifted by $n/R$ where $R$ is the radius
of the extra dimension. In addition,  each non-zero KK
mode level contains 3 additional Higgs-like particles.\footnote{The Higgs
and the 5-modes of the gauge fields provide 8 degrees of freedom, a
combination of which provide the 4 longitudinal modes for the gauge
bosons, leaving 4 Higgs-like states.}  Due to the compactification on $S^1/Z_2$, the KK-levels generically
mix at loop level, however a $Z_2$ parity is left, preventing mixing
between even and odd KK-levels, guaranteeing stability of the LKP.
Matching the zero modes to the
SM fields leaves only the compactification radius $R$
and the Higgs mass $m_H$ as free parameters.

To quantize this non-renormalizable theory it has to be considered as
an effective field theory. Naive dimensional analysis implies a cutoff
around the $50^{\mbox{th}}$ KK mode. At loop level, KK-number violating interactions are induced, whose
counter terms are given by boundary localized kinetic terms. Renormalization only determines their divergent part,
leaving their finite part as a free parameter. The presence of those
brane localized terms changes the boundary conditions for the 5D
fields and therefore has an impact on the mass spectrum. In the ``Minimal UED model''(MUED) these new free
parameters are chosen such that they all vanish at the cutoff scale. However, by running
them down to the electroweak scale, they still affect the mass
spectrum, lifting the approximate mass degeneracy at the first (and
higher) KK-levels and providing an MSSM-like spectrum.\cite{cheng}  

\section{Constraints from electroweak precision measurements and implications for KK dark matter}\label{s2}
The most obvious constraint on UED models arises from lack of production of KK-excitations. KK-excitations
also contribute in loop-corrections and thereby affect precision measurements. Appelquist {\it et al.}\cite{appelquist} determined the beyond standard model contribution to the gauge boson self-energies from which the  
electroweak precision parameters $S,T,U$ follow. Comparison with the experimental data yields a constraint of $1/R> 300$  GeV.

Recently, by including LEP2 data, an extended 
set of electroweak precision parameters $\hat{S},\hat{T},\hat{U}, W,X,Y$ has been determined,\cite{barbieri} all of which are defined in terms of beyond standard
model contributions to the gauge boson self-energies.

\begin{figure}
\begin{center}
\newlength{\picwidth}
\setlength{\picwidth}{2.4in}
\newlength{\picwidthh}
\setlength{\picwidthh}{2.5in}
\begin{tabular}{cc}
	%\hline
	\makebox(200,180)[c]
		{\parbox[b]{\picwidth}
		{\resizebox{\picwidth}{!}{\includegraphics{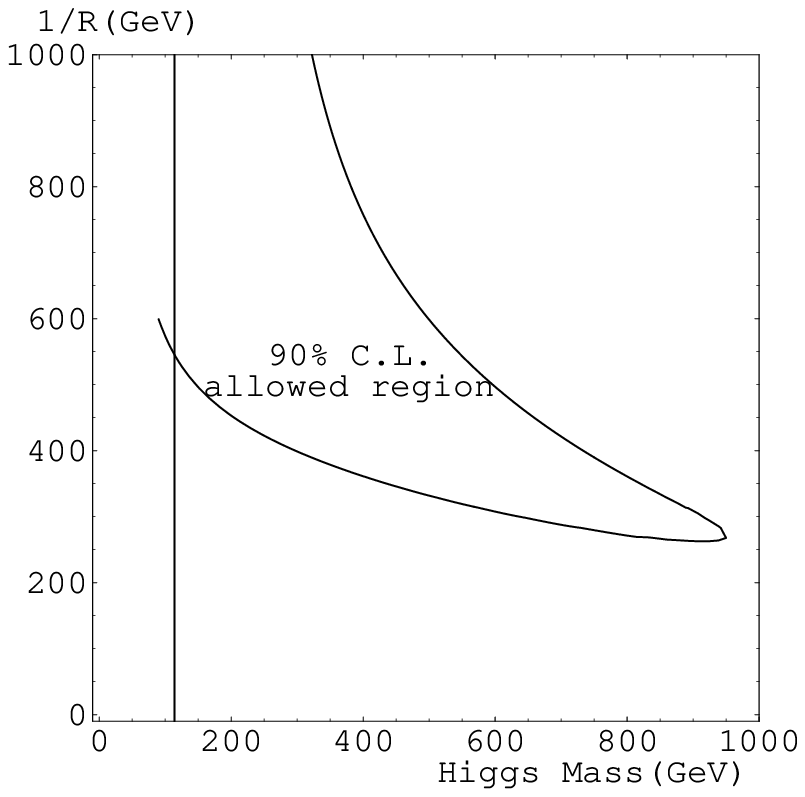}}}}&
	\makebox(200,180)[l]
		{\parbox[t]{\picwidthh}
		{\vspace{10pt}
		\psfrag{mH}[c]{$m_H$ [GeV]}
		\psfrag{Rinv}[c]{\parbox{10pt}{ $R^{-1}$\\ \mbox{[GeV]}}}
		{\resizebox{\picwidthh}{2.5in}{\includegraphics{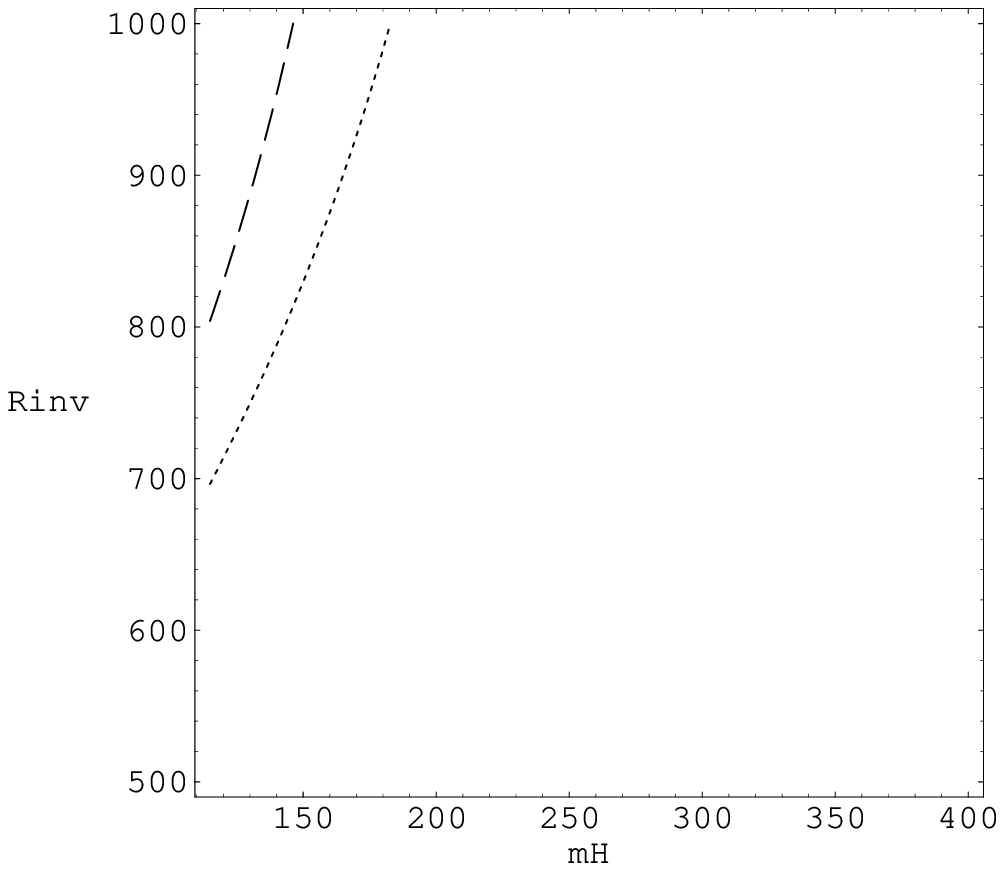}}}}}\\
	%\hline
	\makebox(200,10)[t]
	{\parbox[tl]{200pt}
		{\tiny (a) From Appelquist {\it et al.}\cite{appelquist}: UED parameter space before LEP2.}}&
	\makebox(200,10)[t]
	{\parbox[tl]{200pt}
		{\tiny (b)  Parameter space after inclusion of LEP2 data.\cite{flacke2}}}%\\
%\hline
\end{tabular}
\end{center}
\caption{Constraints on the UED parameter space due to elctroweak precision data.}
\label{Fig1}
\end{figure}

We extend the analysis of Appelquist {\it et al.}, incorporating the full $\hat{S},\hat{T},\hat{U}, W,X,Y$ parameter set.\cite{flacke2} We furthermore incorporate a
fit to the full 2-loop standard model corrections, provided Barbieri
{\it et al.}\cite{barbieri} Our result for the $\chi^2$-fit to the
$\hat{S},\hat{T},\hat{U}, W,X,Y$ parameters is shown in
Fig.~\ref{Fig1}~(b).
The constraint on the size of the extra dimension is improved to $R^{-1} >
700$ (800) GeV at the 99\% (95\%) confidence level. 

Turning to cosmology, the dark matter candidate in UED models provides
an independent constraint on the compactification radius. In the MUED
model, the LKP is the first KK excitation of the $U(1)_Y$ gauge
boson, $B^{(1)}$. Assuming decoupling in thermal equilibrium, its number density
evolves as
\begin{equation}
\frac{dn_{B^{(1)}}}{dt} + 3 H n_{B^{(1)}}
= -<\sigma v> \bigg[(n_{B^{(1)}})^2 - (n^{\rm{eq}}_{B^{(1)}})^2 \bigg], 
\end{equation}
where $H$ is the Hubble rate, $n^{\rm{eq}}_{B^{(1)}}$ denotes the equilibrium number
density of the LKP, and $<\sigma v>$ is the LKP's self-annihilation cross section.
Solving the Boltzmann equation yields the relic density for cold dark
matter measured by WMAP, provided that $m_{B^{(1)}}\sim 1$ TeV.
The above treatment does not take coannihilation with other KK
particles into account, which can considerably change the LKP relic density
in dependence of the exact mass spectrum. The LKP relic density for the
mass spectrum of the MUED has been calculated.\cite{matchev} The
compactification radius leading to the observed relic density is 500
GeV $<1/R<$ 600 GeV.
This value shows tension with the electroweak precision constraint of
$1/R>$ 700 GeV and points towards the necessity to include boundary terms. These would in
particular modify the relic density which strongly depends on the detailed mass spectrum.

\section{Magnetic dipole moment of Kaluza-Klein neutrino dark matter}\label{s3}
If non-minimal boundary terms are taken into account, it is not
guaranteed anymore that the LKP is given by the $B^{(1)}$. Another
WIMP KK-mode and potential candidate is the first KK-mode neutrino, $\nu^{(1)}$. A strong lower bound of
$1/R\gtrsim 50$ TeV arises from direct detection when considering
tree level $Z$ exchange with nuclei.\cite{servant}

\begin{figure}
\begin{center}
\newlength{\picwidtha}
\setlength{\picwidtha}{1.2in}
\newlength{\picwidthaa}
\setlength{\picwidthaa}{3.5in}
\begin{tabular}{cc}
	%\hline
	\makebox(180,190)[l]
		{\parbox[b]{\picwidtha}
		{\resizebox{\picwidtha}{!}{\includegraphics[height=70pt,width=70pt]{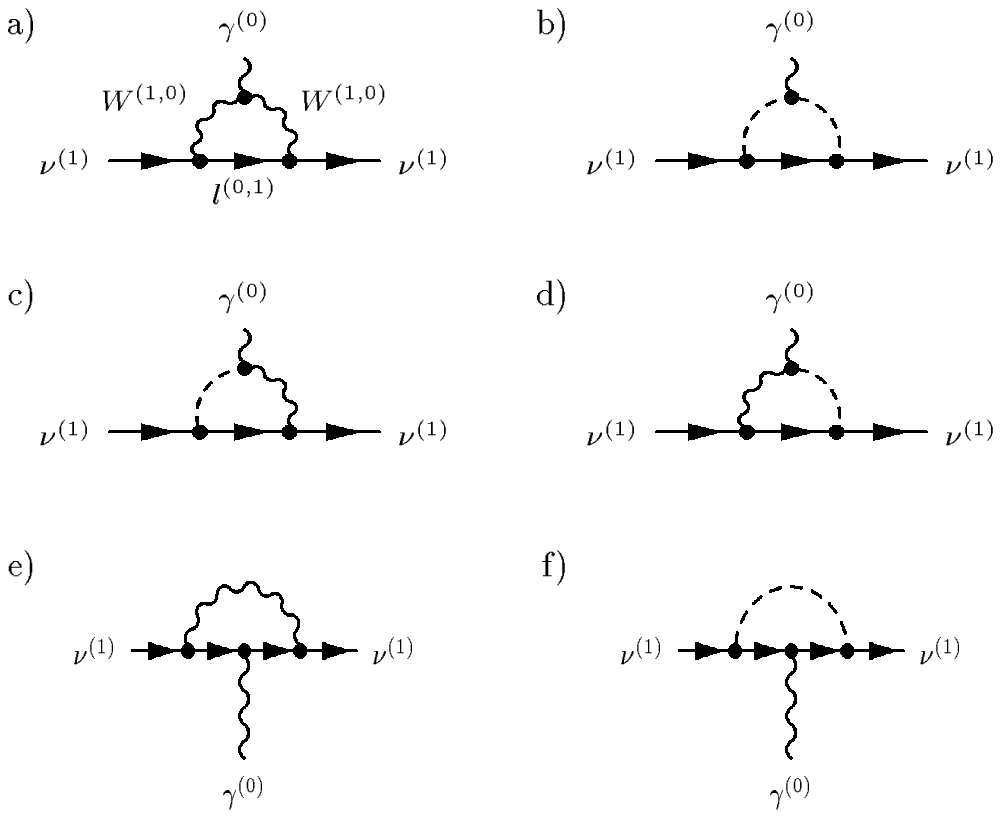}}}}&
	\makebox(250,190)[r]
		{\parbox[t]{\picwidthaa}
		{\vspace{10pt}
		\psfrag{mH}[c]{$m_H$ [GeV]}
		\psfrag{Rinv}[c]{\parbox{10pt}{ $R^{-1}$\\ \mbox{[GeV]}}}
		{\resizebox{\picwidthaa}{2.5in}{\includegraphics{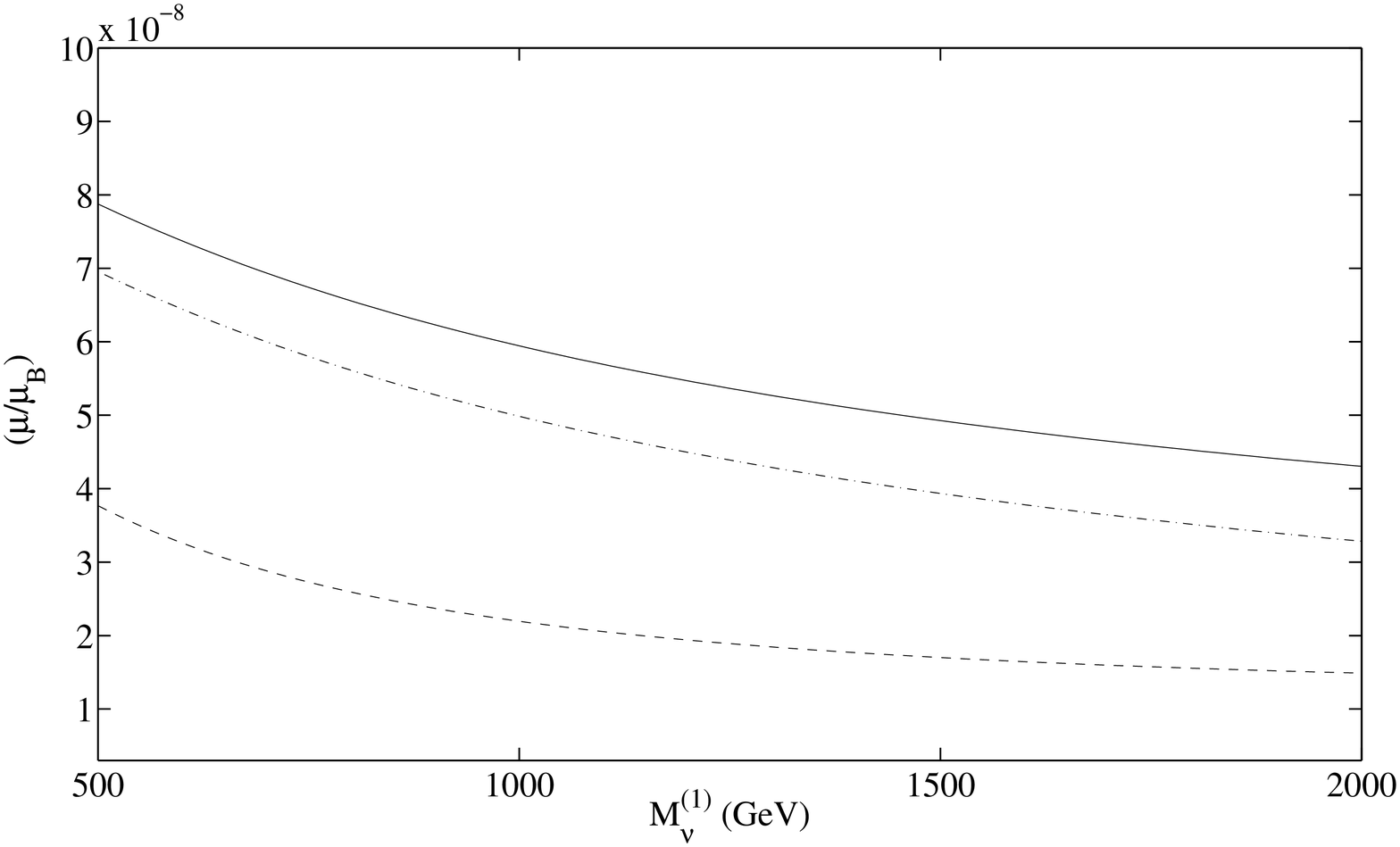}}}}}\\
	%\hline
	\makebox(180,40)[t]
	{\parbox[tl]{150pt}
		{\tiny (a) One-loop contributions to the $\nu^{(1)}$ magnetic dipole moment.}}&
	\makebox(250,40)[t]
	{\parbox[tl]{230pt}
		{\tiny (b) $\nu^{(1)}$ magnetic moment vs. $M_{\nu^{(1)}}$. \\
		Upper line: $M_{e^{(1)}}$-$M_{\nu^{(1)}}$ degeneracy with tree-level $M_{W^{(1)}}$. \\
		Middle line: 5\% splitting in  $M_{e^{(1)}}$/$M_{\nu^{(1)}}$ with tree-level $M_{e^{(1)}}$. \\
		Lowest line: $M_{e^{(1)}}$-$M_{\nu^{(1)}}$ degeneracy with 5\% split of $M_{W^{(1)}}$/$M_{\nu^{(1)}}$.}}%\\
%\hline
\end{tabular}
\end{center}
\caption{One-loop contributions and results for the $\nu^{(1)}$ magnetic dipole moment.}
\label{Fig2}
\end{figure}

Apart from this process, the $\nu^{(1)}$, being a Dirac fermion, also
has a magnetic dipole moment, induced at loop level. The one-loop graphs contributing to the magnetic dipole moment are
given in Fig. \ref{Fig2} (a). From them, we obtain the semi-analytic result:
\begin{equation} 
\label{dipole_result} 
\mu =   \frac{eg^2}{(4\pi)^2}\frac{1}{M_{\nu^{(1)}}}\left\{\frac{3}{2}\ln (\epsilon)+r+\frac{7}{2}+\frac{1}{2r}-\frac{5}{2}(r-1)\ln \left(\frac{r}{r-1}\right)-(r-1)^2\ln \left(\frac{r}{r-1}\right) + \mathcal{O}(\sqrt{\epsilon})\right\}	
\end{equation}
with the approximation $\epsilon\equiv M_{W^{(0)}}^2/M_{\nu^{(1)}}^2\ll 1$ and $r\equiv M_{W^{(1)}}^2/M_{\nu^{(1)}}^2\simeq 1$.
The numerical results are given in Fig. \ref{Fig2} (b).
For $M_{\nu^{(1)}}\sim $ 1 TeV, the magnetic moment of the KK-neutrino is of the order of
$10^{-7}\mu_B$ - $10^{-8}\mu_B$, with the exact value depending on the mass
splitting between $M_{e^{(1)}}$, $M_{\nu^{(1)}}$ and $M_{W^{(1)}}$. 
From eqn. (\ref{dipole_result}), the magnetic moment scales with
$1/M_{\nu^{(1)}}\equiv R$.
To obtain a constraint on $R$, the magnetic dipole moment is to be
compared with the experimental constraint on the magnetic moment of
fermionic dark matter. Our own very rough estimate of it is
$\mu\lesssim 10^{-8} \mu_B$, indicating that this leads to a  constraint on $R$ in the same ballpark as
direct $Z$ exchange. The only detailed analysis of the experimental constraint known to us is
currently under revision and its final result will give the answer to whether and for which mass spectra direct $Z$ exchange or magnetic dipole moments provide the dominant constraints.

\section{Conclusions}
We presented the currently strongest constraint on the radius of the UED model. It is determined from electroweak precision tests, yielding a bound of $M_c \equiv R^{-1} > 700$ (800) GeV at the 99\% (95\%) confidence level. Comparison with the radius preferred by the relic density of dark
matter in the MUED model (500 GeV $<R^{-1} <$  600 GeV) points towards the necessity of including non-minimal boundary terms which can change the nature of LKP. 

We presented the calculation of the loop-induced dipole magnetic
moment of the KK neutrino and await results on the experimental value
to compare to.

\section*{Acknowledgements}
I would like to thank the organizers of the ``Rencontres de Moriond 2006'' on Electroweak interactions and Unified theories for providing the wonderful background for this stimulating conference. I am grateful to the European Union ``Marie Curie'' programme and to Wolfson College, Oxford, for their financial support.

\end{document}